\documentclass[a4paper]{article}

\usepackage{INTERSPEECH2018}
\usepackage{subcaption}
\usepackage{color,xcolor}
\usepackage{setspace}\usepackage{amsmath}

\graphicspath{{figures/}}

\allowdisplaybreaks[4]
\title{Empirical Evaluation of Speaker Adaptation on DNN based Acoustic Model}
\name{Ke Wang$^{1,2}$, Junbo Zhang$^2$, Yujun Wang$^2$, Lei Xie$^{1*}$\thanks{*Corresponding author}}
\address{
  $^1$Shaanxi Provincial Key Laboratory of Speech and Image Information Processing,\\
  School of Computer Science, Northwestern Polytechnical University, Xi'an, China\\
  $^2$Xiaomi, Beijing, China}
\email{\{kewang, lxie\}@nwpu-aslp.org, \{zhangjunbo, wangyujun\}@xiaomi.com}

\begin{document}
\begin{spacing}{0.90}

\maketitle
\begin{abstract}
Speaker adaptation aims to estimate a speaker specific acoustic model from a speaker independent one to minimize the mismatch between the training and testing conditions arisen from speaker variabilities.  A variety of neural network adaptation methods have been proposed since deep learning models have become the main stream.
But there still lacks an experimental comparison between different methods, especially when DNN-based acoustic models have been advanced greatly. In this paper, we aim to close this gap by providing an empirical evaluation of three typical speaker adaptation methods: LIN, LHUC and KLD. Adaptation experiments, with different size of adaptation data, are conducted on a strong TDNN-LSTM acoustic model. More challengingly, here, the source and target we are concerned with are standard Mandarin speaker model and accented Mandarin speaker model. We compare the performances of different methods and their combinations. Speaker adaptation performance is also examined by speaker's accent degree. 

\end{abstract}

\section{Introduction}

Speech recognition accuracy has been significantly improved since the use of deep learning models (DLMs), or more specifically, deep neural networks (DNNs)~\cite{dahl2012context,hinton2012deep}. Various models, such as convolutional neural networks (CNNs)~\cite{abdel2012applying,abdel2013exploring}, time-delay neural networks (TDNNs)~\cite{peddinti2015time}, long short-term memory (LSTM) recurrent neural networks (RNNs)~\cite{sak2014long,sak2015fast} and their variants~\cite{zhang2016highway,zhang2015feedforward} and combinations~\cite{sainath2015convolutional}, have been developed to further improve the performance. However, the accuracy of an automatic speech recognition (ASR) system in real applications still lags behind that in controlled testing conditions. This raises the old and unsolved problem called \textit{training-testing mismatch}, i.e., the training set cannot match the new acoustic conditions or fails to generalize to new speakers. Thus a variety of acoustic model compensation and adaptation methods have been proposed, to better deal with unseen speakers and mismatched acoustic conditions.

This study specifically focuses on \textit{speaker adaptation}, i.e., modifying a general model, commonly a speaker-independent acoustic model (SI AM), to work better for a specific new speaker, though the same adaptation technique can be applied to other mismatched conditions. The history of acoustic model speaker adaptation can be traced back to the GMM-HMM era~\cite{Woodland2001Speaker,Gauvain1994Maximum,Legetter1995Maximum,Digalakis1995Speaker,Gales1998Maximum,Gales2000Cluster,Kuhn2000Rapid,Uebel1999An}, while the focus has been shifted to neural networks since the rise of DLMs. Various approaches have been developed for neural network acoustic model adaptation~\cite{neto1995speaker,li2010comparison,gemello2007linear,swietojanski2014learning,saon2013speaker,miao2015speaker,abdel2013fast,yu2013kl,senior2014improving,huang2017bayesian} and they can be roughly categorized into three classes: speaker-adapted layer insertion, subspace method and direct model adapting.

%

In the category of speaker-adapted layer insertion, linear transformation, which augments the original network with certain speaker-specific linear layer(s), is a simple-but-effective approach. Common methods include linear input network (LIN)~\cite{neto1995speaker,li2010comparison}, linear hidden network (LHN)~\cite{gemello2007linear}, and linear output network (LOH)~\cite{li2010comparison}, just to name a few. Among them, LIN is the most popular one. Learning hidden unit contribution (LHUC)~\cite{swietojanski2014learning} is another type of speaker-adapted layer insertion method that makes the SI network parameters to be speaker-specific by inserting special layers to control the amplitude of the hidden layers. 

Another category, subspace method, aims to find a low dimensional speaker subspace that is used for adaptation. The most straightforward application is to use subspace-based features, e.g., i-vectors~\cite{saon2013speaker,miao2015speaker}, as a supplement of acoustic features in the neural network for acoustic model training, or speaker adaptive training (SAT). Another approach, serving the same purpose with auxiliary features, is called speaker codes~\cite{abdel2013fast}. A specific set of network units for each speaker is connected and optimized with the original SI network. Note that i-vector based SAT has become a standard in the training of deep neural network acoustic models~\cite{peddinti2015time,miao2015speaker,senior2014improving,saon2015ibm,povey2016purely,xiong2017microsoft} as this simple trick can bring small-but-consistent improvement.

A straightforward idea is to use new speaker's data to adapt the DNN parameters directly. Retraining/fine-tuning the SI model using the new data is the simplest way, which is also called retrained speaker independent (RSI) adaptation~\cite{neto1995speaker}. To avoid over-fitting, conservative training, such as Kullback-Leibler divergence (KLD) regularization~\cite{yu2013kl} is further introduced. This approach tries to force the posterior distribution of the adapted model to be closer to that estimated from the SI model, by adding a KLD regularization term to the original cross entropy cost function to update the network parameters. Although quite effective, this approach results in an individual neural network for each speaker.

To the best of our knowledge, there still lacks a thorough experimental comparison between different speaker adaptation methods in the literature, especially when the DNN-based acoustic models (AMs) have been advanced greatly since the introduction of these adaptation techniques. In this paper, we aim to close this gap by providing an empirical evaluation of three typical speaker adaptation methods: LIN, LHUC and KLD. Adaptation experiments are conducted on a strong TDNN-LSTM acoustic model (well trained i-vector based SAT-DNN acoustic model with cMLLR~\cite{Legetter1995Maximum,Gales1998Maximum}) tested with different size of adaptation data. More challengingly, here, the source and target we are concerned with are standard Mandarin speaker model and accented Mandarin speaker model. We compare the performance of different methods and their combinations. The speaker adaptation performance is also examined by speaker's accent degree. In a word, we would like to provide readers a big picture on the selection of speaker adaptation techniques.

%

The rest of this paper is organized as follows. In Section~\ref{section:algorithms}, we briefly introduce LIN, KLD, LHUC and give a discussion on their abilities. Next, we describe a series of experiments and report the results in Section~\ref{section:experiments}. Finally, some conclusions are drawn in Section~\ref{section:conclusions}.

\section{Speaker adaptation algorithms}\label{section:algorithms}

\subsection{LIN}

\begin{figure}[tbp]
  \centering{}
  \includegraphics[width=0.8\linewidth]{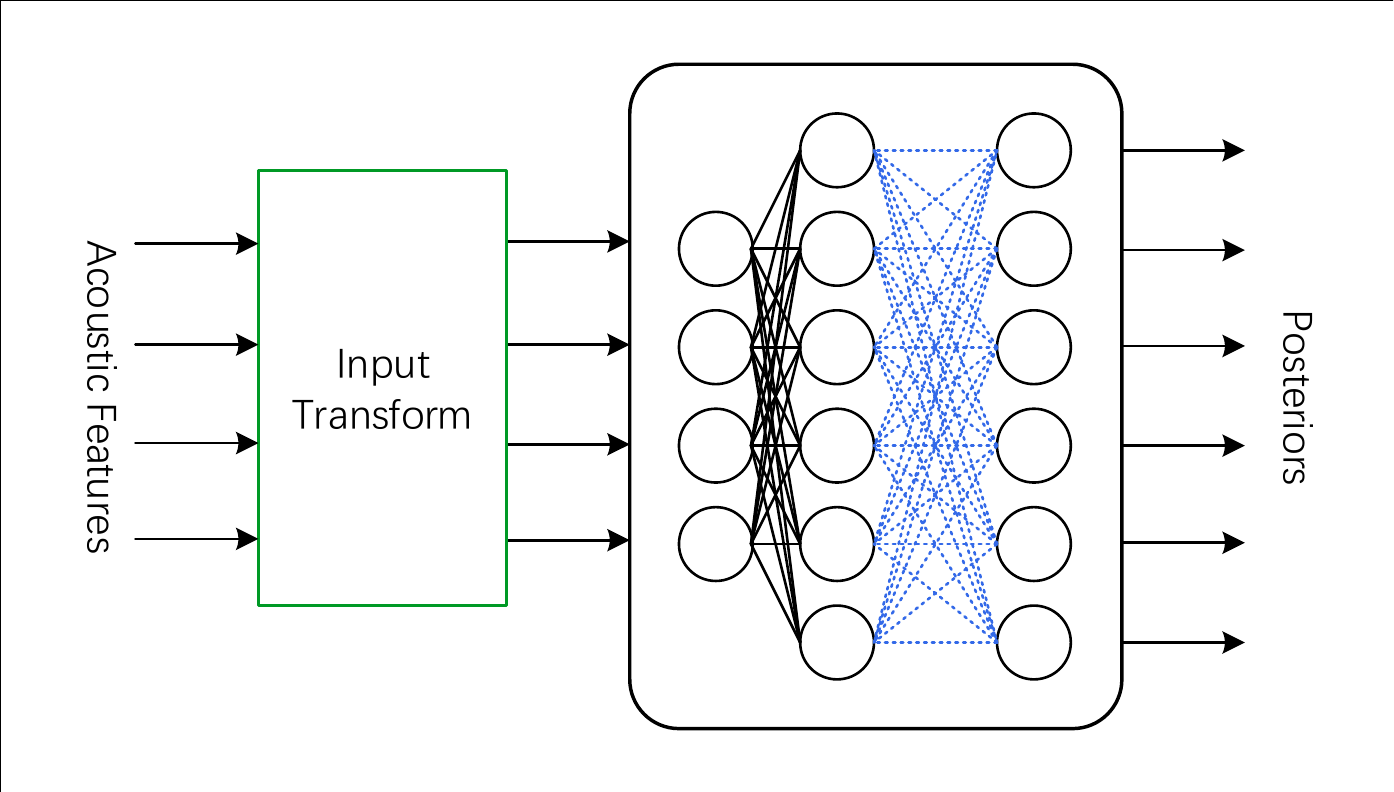}
  \caption{Linear input network.}
  \label{fig:lin}
  \vspace{-0.6cm}
\end{figure}

Linear input network (LIN)~\cite{neto1995speaker,li2010comparison} is a classical input transformation approach for neural network adaptation. As shown in Figure~\ref{fig:lin}, LIN assumes that the mismatch between training and testing can be captured in the feature space by employing a trainable linear input layer which maps speaker dependent speech to speaker independent network (i.e., acoustic model). The inserted layer usually has the same dimension as the original input layer and is initialized to an identity weight matrix and 0 bias. Unlike other layers of the neural network, linear activation function $f(x)=x$ is used for this additional layer.

During adaptation, standard error back-propagation (BP) is used to update the LIN's parameters while keeping all other network parameters fixed, by minimizing the loss function (e.g., cross entropy, mean square error) of the original AM. After adaptation, each speaker-specific LIN captures the relations between the speaker and the training space. Finally, for each testing speaker, the corresponding LIN is selected to do feature transformation and the transformed vector is directly fed to the original unadapted AM for speech recognition.

\subsection{KLD Regularization}


As a popular conservative training adaptation technique, Kullback-Leibler divergence (KLD)~\cite{yu2013kl} regularization tries to force the posterior distribution of the adapted model to be closer to that estimated from the SI model. By contrast, the $L_2$ regularization aims to keep the parameters of adapted model to be closer to those of the SI model.

For acoustic model training, it is typical to minimize the cross entropy (CE)
\begin{footnotesize}
\begin{equation}\label{eq:CE}
  \mathcal{F}_{CE} = - \frac{1}{N} \sum_{t=1}^{N} \sum_{y=1}^{S} \tilde{p}(y|{x_t}) \log p(y|{x_t})\,,
\end{equation}
\end{footnotesize}where $N$ is the number of training samples, $S$ is the total number of states, $\tilde{p}(y|x_t)$ is the target probability and $p(y|x_t)$ is neural network's output posteriors. We usually use a hard alignment from an existing ASR system as the training labels and set $\tilde{p}(y|x_t)=\delta(y=s_t)$, where $\delta$ is the Kronecker delta function and $s_t$ is the label of $t$-th sample. By adding the KLD term to Eq.~(\ref{eq:CE}) we get the following optimization criterion:
\begin{footnotesize}
\begin{align}\label{eq:KLD_1}
 \widehat{\mathcal{F}}_{CE} &= (1-\rho) \mathcal{F}_{CE} - \rho \frac{1}{N} \sum_{t=1}^{N} \sum_{y=1}^{S}p^{SI}(y|x_t) \log p(y|x_t) \notag \\[-0.2cm] 
                            &= - \frac{1}{N} \sum_{t=1}^{N} \sum_{y=1}^{S} \left[ (1-\rho) \tilde{p}(y|x_t) + \rho p^{SI}(y|x_t) \right] \log p(y|x_t) \notag \\[-0.1cm]
                            &= - \frac{1}{N} \sum_{t=1}^{N} \sum_{y=1}^{S} \hat{p}(y|x_t) \log p(y|x_t)\,,
\end{align}
\end{footnotesize}where $\rho$ is regularization weight and we have defined
\begin{footnotesize}
\begin{equation}\label{eq:KLD_2}
  \hat{p}(y|x_t) \triangleq (1-\rho) \tilde{p}(y|x_t) + \rho p^{SI}(y|x_t)\,.
\end{equation}
\end{footnotesize}

By comparing Eq.~(\ref{eq:CE}) and Eq.~(\ref{eq:KLD_1}), we can find that applying KLD is equivalent to changing the target distribution in the conventional BP algorithm. When $\rho=0$, we can regard this configuration as RSI, i.e., retraining the SI model directly using the traditional CE loss.

\subsection{LHUC}

\begin{figure}[tbp]
  \centering{}
  \includegraphics[width=0.8\linewidth]{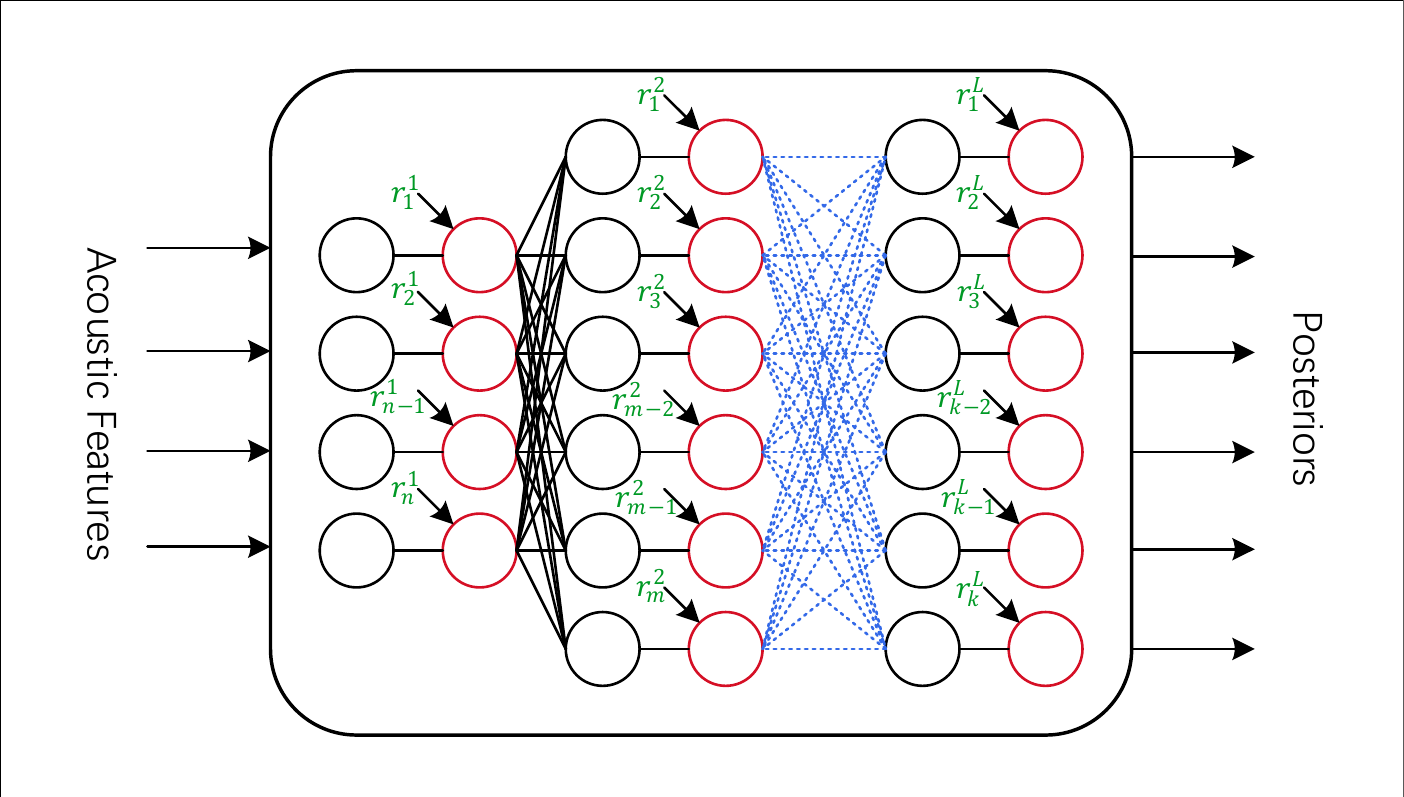}
  \caption{Learning hidden unit contribution.}
  \label{fig:lhuc}
  \vspace{-0.5cm}
\end{figure}

As shown in Figure~\ref{fig:lhuc}, learning hidden unit contribution (LHUC)~\cite{swietojanski2014learning} modifies the SI model by defining a set of speaker dependent parameters $\bm{\theta}$ for a specific speaker, where $\bm{\theta}=\left\{\bm{r}^1,\cdots,\bm{r}^L\right\}$ and $\bm{r}^l$ is the vector of speaker dependent parameters for $l$-th hidden layer. Then the element-wise function $a(\cdot)$ is adopted to constrain the range of $\bm{r}^l$ and the speaker dependent hidden layer output can be defined as the following function:
\begin{footnotesize}
\begin{equation}\label{eq:LHUC}
  \bm{h}^l = a(\bm{r}^l) \circ \phi^l(\bm{W}^{l\top}\bm{h}^{l-1})\,,
\end{equation}
\end{footnotesize}where $\circ$ is an element-wise multiplication and $a(\cdot)$ is typically defined as a sigmoid with amplitude 2, i.e.,

\begin{footnotesize}
\begin{equation}\label{eq:LHUC_a}
  a(\bm{r}^l) \triangleq \frac{2}{1+\exp(-\bm{r}^l)}\,,
\end{equation}
\end{footnotesize}to constrain the range of $\bm{r}$'s elements to $\left[0,2 \right]$.

LHUC, given adaptation data, actually rescales the contributions (amplitudes) of the hidden units in the model without actually modifying their feature receptors.  At the training stage, $\bm{\theta}$ is optimized with the standard BP algorithm while keeping all the other parameters fixed for a specific speaker. During the testing stage, the corresponding $\bm{\theta}$ is chosen to constrain the amplitudes of hidden units in order to get more accurate posterior probability for the speaker.

\subsection{Discussion and Combination}

\begin{footnotesize}
\begin{table*}[tbp]
  \caption{CERs of each speaker on baseline TDNN-LSTM i-vector based acoustic model. ``S'', ``M'' and ``H'' are short forms for ``slight'', ``medium'' and ``heavy'' separately.}
  \label{tab:baseline}
  \centering
    \begin{tabular}{*{12}{c}}
      \toprule
      \textbf{Speaker} & \textbf{S01} & \textbf{S02}  & \textbf{S03} & \textbf{S04} & \textbf{S05} & \textbf{S06} & \textbf{S07}  & \textbf{S08} & \textbf{S09} & \textbf{S10} & \textbf{Avg} \\
      \hline
      \textbf{Accent}   & S      & M       &  S     & M       & H        & H       &  H      & M       & H       &  M      &  -      \\
      \textbf{CER (\%)} & $3.00$ & $21.63$ & $9.09$ & $16.40$ & $56.62$  & $40.07$ & $36.61$ & $11.16$ & $31.74$ & $22.28$ & $24.86$ \\
      \bottomrule
    \end{tabular}
\end{table*}
\end{footnotesize}

\begin{figure*}[tbp] 
\centering
\begin{subfigure}[t]{0.33\textwidth}
  \centering
  \includegraphics[width=1.0\textwidth]{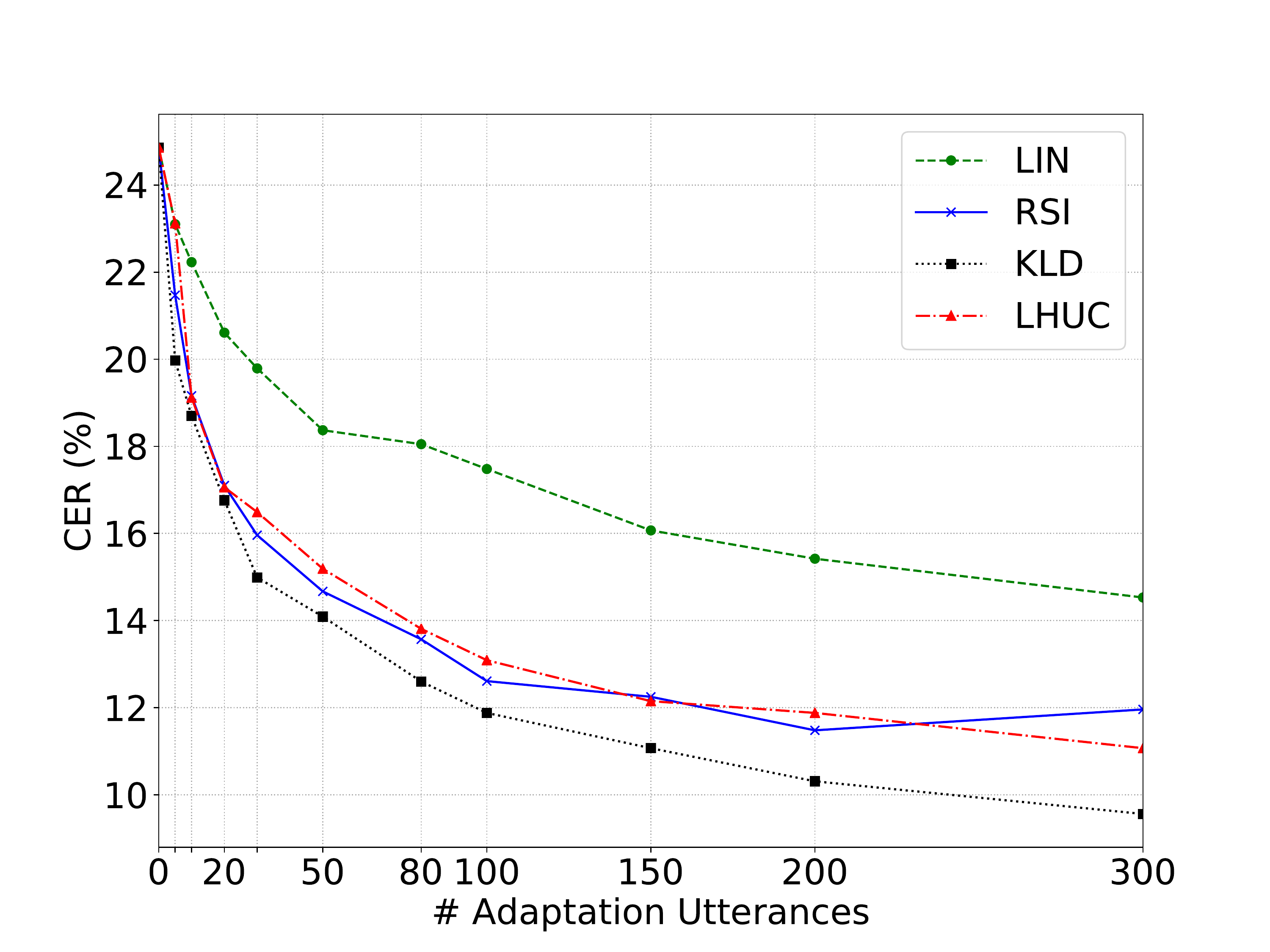}
  \subcaption{}
  \label{fig:summary}
\end{subfigure}
\begin{subfigure}[t]{0.33\textwidth}
  \centering
  \includegraphics[width=1.0\textwidth]{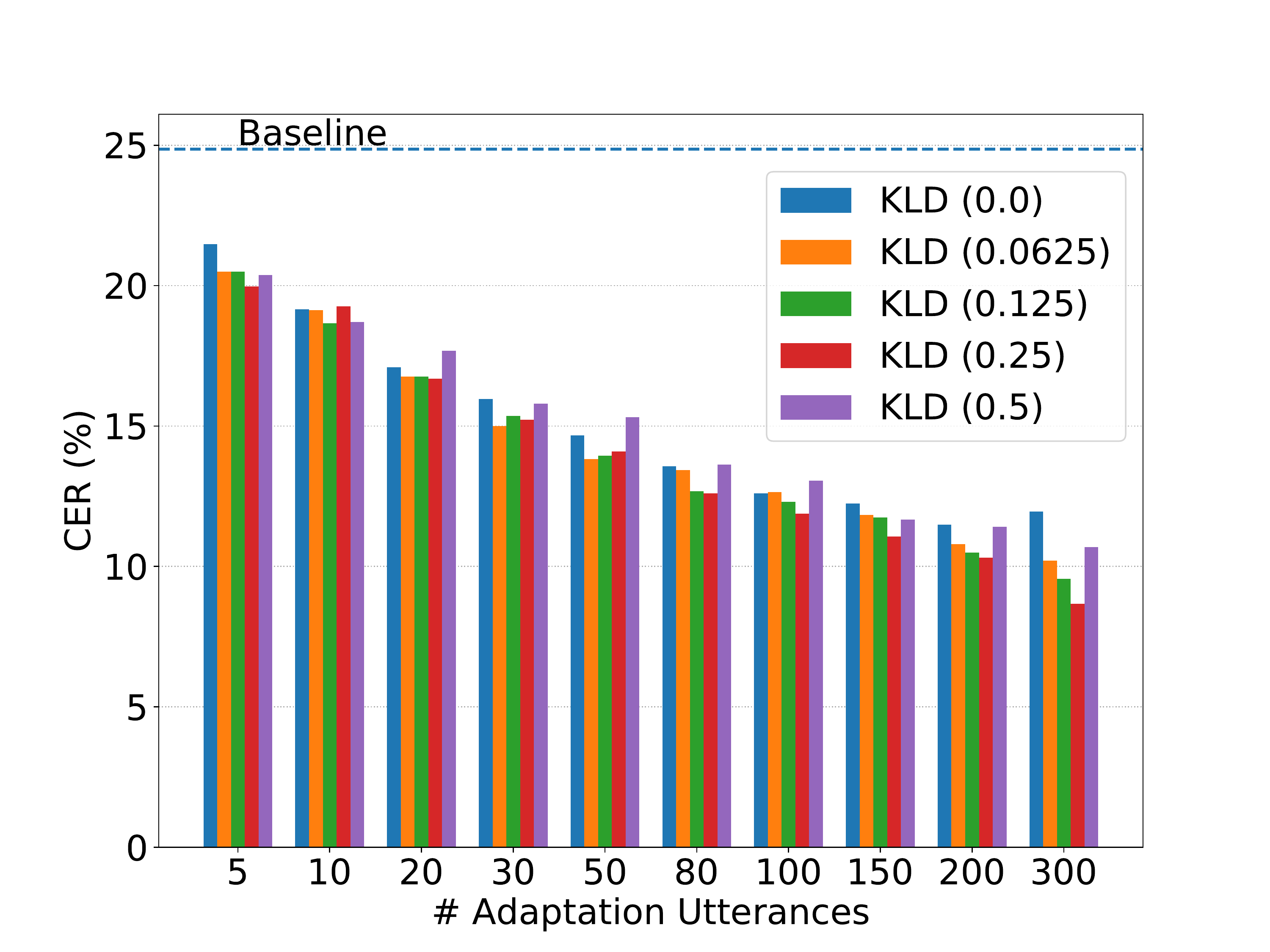}
  \subcaption{}
  \label{fig:bar_kld}
\end{subfigure}
\begin{subfigure}[t]{0.33\textwidth}
  \centering
  \includegraphics[width=1.0\textwidth]{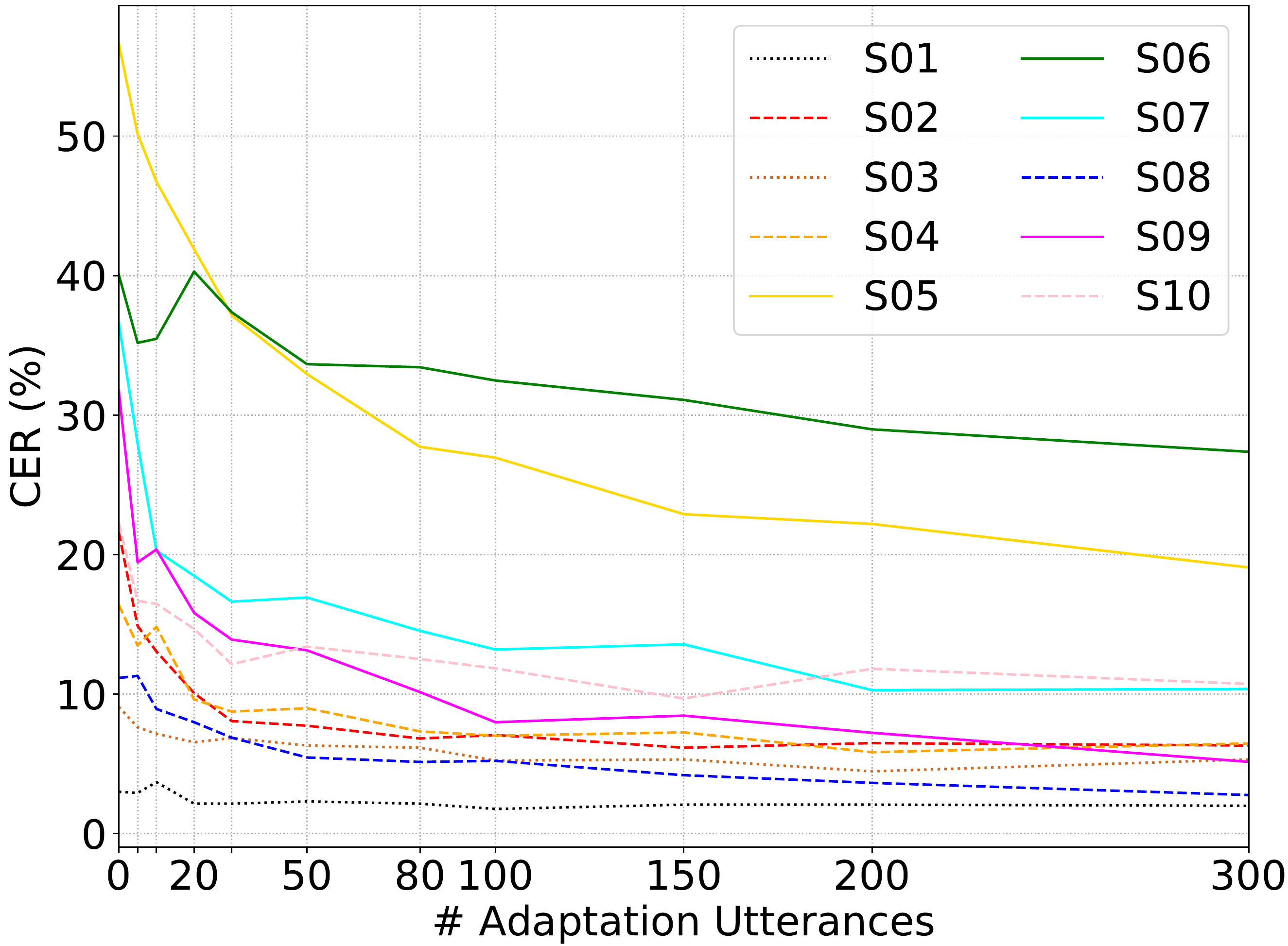}
  \subcaption{}
  \label{fig:each_kld}
\end{subfigure}
\caption{CERs (\%) for different amount of adaptation data. (a) Comparison of different adaptation methods. (b) KLD adaptation with different regularization weights $\rho$. The dashed line is the baseline CER. (c) KLD adaptation for each speakers.}
\label{fig:summary_families}
\vspace{-0.5cm}
\end{figure*}

We compare the three speaker adaptation approaches in terms of adapted parameter size and modification on the AM.
\begin{itemize}
\item \textbf{Size of Adapted Parameters:} LHUC has minimal adapted parameters, followed by LIN. For KLD regularization, since each speaker has a fully adapted neural network AM, it results in the largest size of adapted parameters.
\item  \textbf{Modification on AM:} In the KLD regularization based adaptation, we do not need to change the original AM network structure, while only changing the loss function. By contrast, we need to adjust the network structure, e.g., inserting layers in the use of LIN and LHUC. However, we need to take extra burden to find an appropriate regularization weight $\rho$ in the KLD regularization based adaptation, which is searched through the validation set.
\end{itemize}

The three approaches perform network adaptation from different aspects and thus can be integrated to expect some extra benefits. LIN and LHUC, both without changing the parameters of the original SI network, can be directly integrated. In other words, LIN's parameters and the speaker dependent parameters $\bm{\theta}$ are updated using the target speaker's data while keeping the parameters of the original network intact. As the KLD adaptation itself needs to update the parameters of the original network, in the integration of LIN/LHUC with KLD, we only use Eq.~(\ref{eq:KLD_1}) as the loss function to update LIN's parameters or/and the speaker dependent parameters $\bm{\theta}$ while still keeping the original SI network parameters unchanged.


\section{Experiments}\label{section:experiments}

\begin{figure}[tbp]
  \centering{}
  \includegraphics[width=0.9\linewidth]{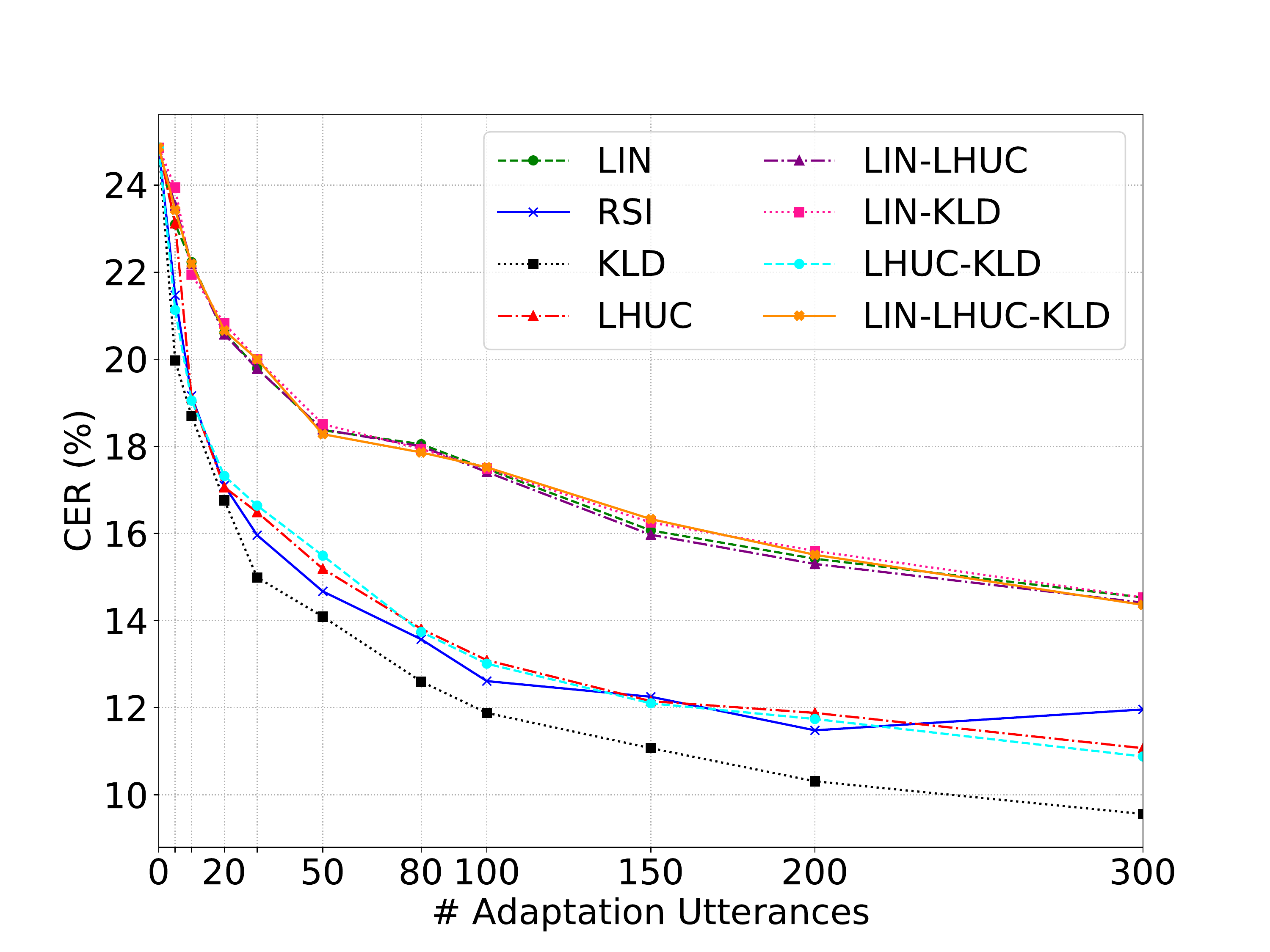}
  \caption{CERs (\%) for different method combinations.}
  \label{fig:summary_combines}
  \vspace{-0.5cm}
\end{figure}

\begin{figure*}[tbp]
\centering
\begin{subfigure}[t]{0.33\textwidth}
  \centering
  \includegraphics[width=1.0\textwidth]{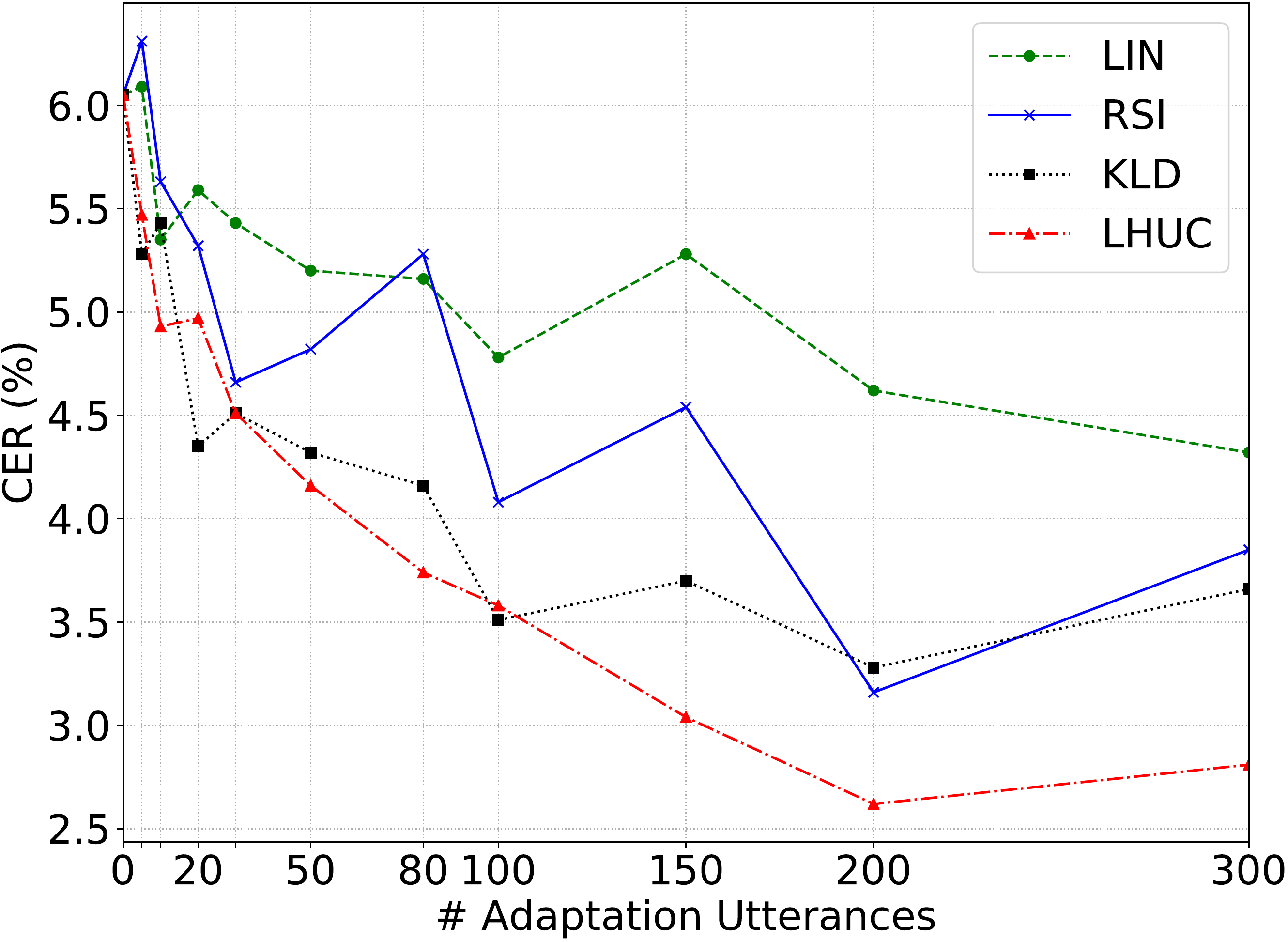} 
  \subcaption{Slight Accent}
  \label{fig:accent_slight}
\end{subfigure}
\begin{subfigure}[t]{0.33\textwidth}
  \centering
  \includegraphics[width=1.0\textwidth]{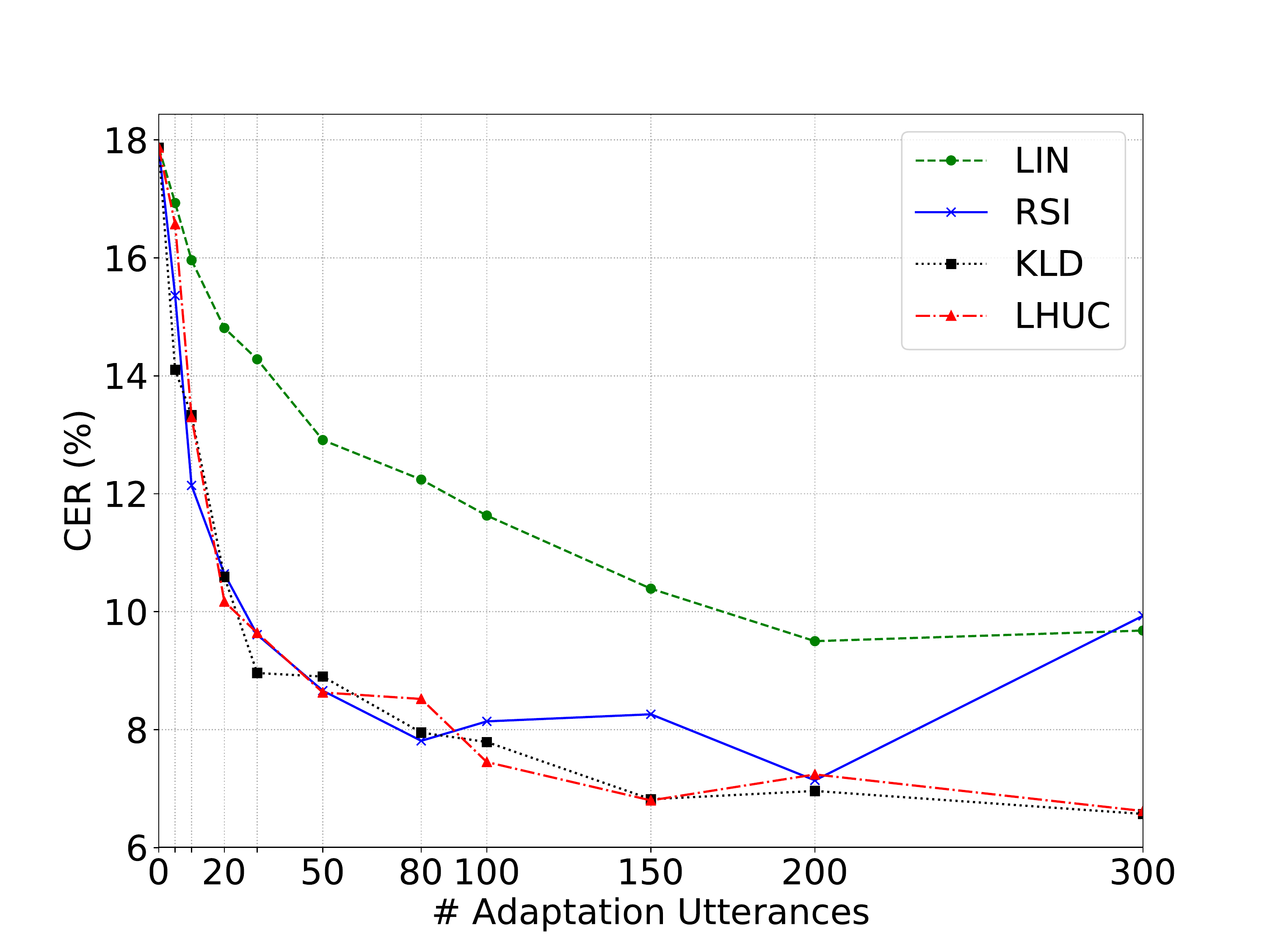}
  \subcaption{Medium Accent}
  \label{fig:accent_moderate}
\end{subfigure}
\begin{subfigure}[t]{0.33\textwidth}
  \centering
  \includegraphics[width=1.0\textwidth]{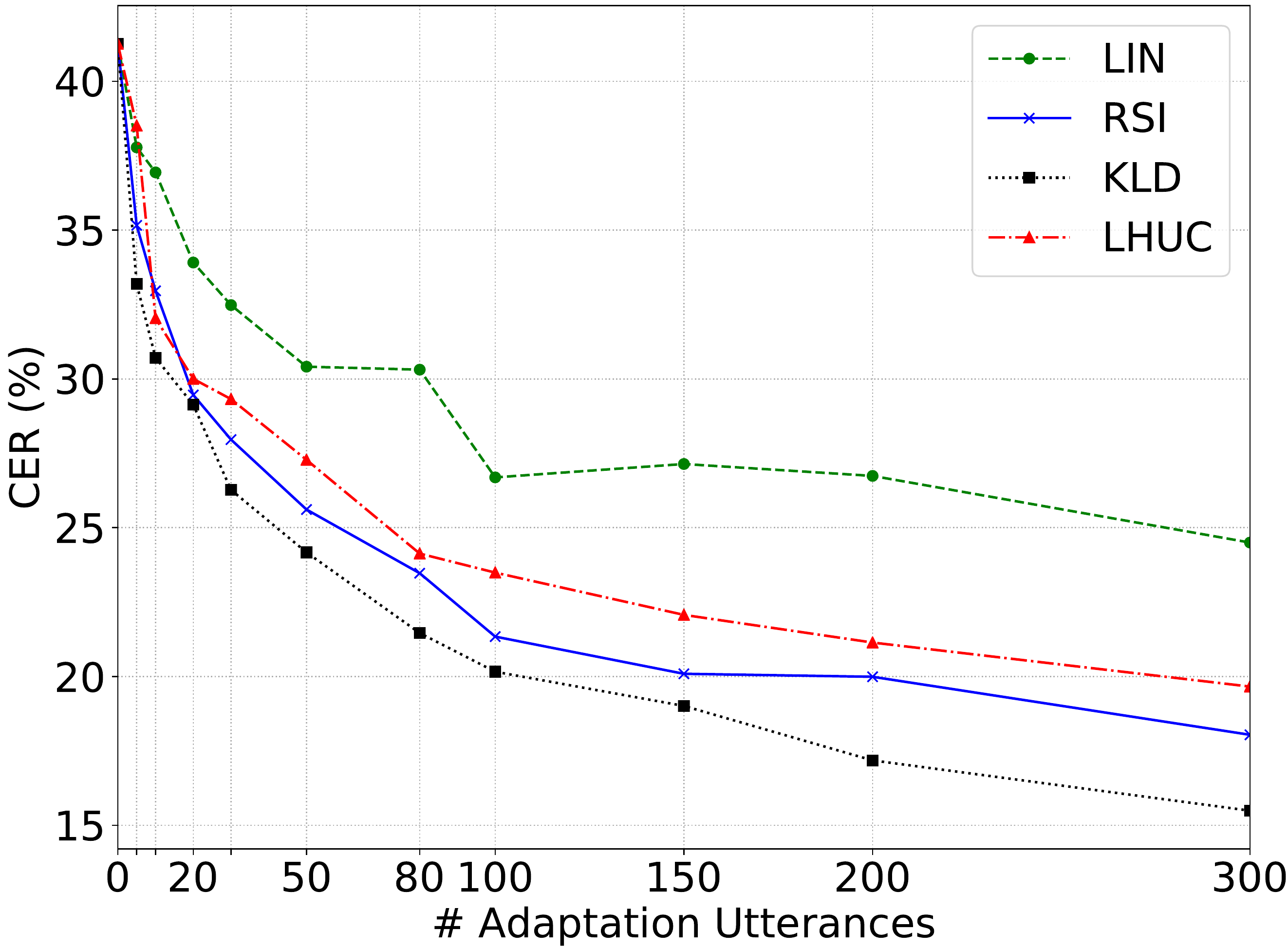}
  \subcaption{Heavy Accent}
  \label{fig:accent_thick}
\end{subfigure}
\caption{CERs (\%) for different methods on different degrees of accent.}
\label{fig:differnet_accent}
\vspace{-0.4cm}
\end{figure*}

\subsection{Experimental setup}

In the experiments, we used a Mandarin corpus that consists of 3,000 speaker (about 1000hrs) with standard accent to build a baseline TDNN-LSTM AM. Before NN model training, the alignments were achieved from a GMM-HMM AM, combined with fMLLR,  trained using the same dataset. Our speaker adaptation dataset consists of 10 Mandarin speakers from Hubei Province of China and each speaker contributes 450 utterances (about 0.5hr/speaker). Note that the 10 speakers have different levels of accents and we expect that a good speaker adaptation technique should handle different levels of accents. We randomly selected 50 utterances as the cross validation set, 100 utterances as the test set and the others as the training set. In the adaptation experiments, we varied the number of training utterances from 5 to 300 to observe the performances of different data size.

For the baseline SI acoustic model, 40-dimensional Mel filter-bank cepstral coefficients (MFCCs) spliced with 2 left, 2 right frames and 100-dimensional i-vector, further transformed to 300-dimension with linear discriminate analysis (LDA), were used as the network input. The output softmax layer has 5,795 units representing senones. Moreover, the TDNN-LSTM model has 6 TDNN layers (520 neurons) and 3 LSTMP layers~\cite{sak2014long} (520 cells with 130 recurrent nodes and 130 non-recurrent nodes). Network training started from an initial learning rate of 0.0003~\footnote{More details about this architecture can be found in Kaldi: egs/wsj/s5/local/nnet3/run\_tdnn\_lstm.sh}. A trigram language model (LM) was used in evaluating both the baseline and the adapted models. Moreover, all of our experiments were based on Kaldi~\cite{povey2011kaldi}.

\subsection{Results of Baseline Model}

Table~\ref{tab:baseline} shows the character error rate (CER) for each speakers, tested with the baseline AM. We can see that the baseline model performs differently for each speaker and the average CER is 24.86\%. The CER has a wide range from 3\% to 56.62\%. We manually checked the recordings from different speakers and found that speaker S05 had heavy accent and speaker S01 had slight Mandarin accent. This huge difference gives the speaker adaptation methods a big challenge. We will report the adaptation results in terms of accent levels later in Section~\ref{sec:diff_accent}. 

\subsection{Comparison of LIN, RSI, KLD and LHUC}

We investigated the adaptation ability of LIN, RSI, KLD and LHUC using different amount of adaptation data. Previous studies on LHUC~\cite{swietojanski2014learning} have demonstrated that adapting more layers in the network can get continuously better accuracy. Hence we inserted LHUC parameters after each hidden layers. For LIN, models were adapted with a small learning rate of 0.00001, while 0.001 and 0.01 were used as an initial learning rate for KLD and LHUC, respectively. From the results shown in Figure~\ref{fig:summary}, we can see that KLD achieves the best performance and is more stable than RSI on different amount of adaptation data for all speakers. LIN, as simple layer-insertion method, is also helpful, but its performance is not as good as the other two. For RSI and LHUC, their performances are comparable in most cases, but over-fitting is occurred for RSI when the adaptation data size exceeds 200.

Furthermore, similar with~\cite{yu2013kl}, we gave a deep investigation on KLD-based adaptation and results are shown in Figure~\ref{fig:bar_kld}. First, unlike the results in~\cite{yu2013kl}, where using small amount of data (5 or 10 utterances) for KLD adaptation is unfortunately harmful, we still can obtain apparent CER reduction when the same size of data are used for adaptation. We believe that this is because our testing speakers have noticeable accents, i.e., the difference between the SI data and the target speaker data is significant. The comparison of different $\rho$ in the range of $[0.0625, 0.5]$ also indicates that reasonable CER reduction can be obtained even with a small $\rho$ for different size of adaptation data.  The figure also clearly shows that a medium regularization weight (e.g., 0.25) is preferred for larger and smaller adaptation sets and a smaller regularization weight (e.g., 0.0625) is better used for medium size of adaptation set. We also compared the performances between different speakers. Results from Figure~\ref{fig:each_kld} shows that KLD works for every testing speaker and the speaker with highest CER on the SI model ( i.e., S5, with the heaviest accent) achieves the largest CER reduction. But with the increase of adaptation data, the gain on each speaker becomes smaller and smaller.

\subsection{Combinations}

We further experimented on method combinations and results are summarized in Figure~\ref{fig:summary_combines}. We can see the combinations of different methods cannot bring salient improvements and the best performance is achieved by KLD only. Even badly, any combination with LIN will drag the performance to LIN.  Combining LHUC with KLD can obtain slightly better result than the vanilla LHUC for very small (less than 10) and large (more than 200) adaptation dataset. But for small adaptation data size (20$\sim$80), LHUC itself performs better.

%
%

\subsection{Different degrees of accent}\label{sec:diff_accent}

As shown in Table~\ref{tab:baseline} earlier, the baseline AM's performance varies on different speakers. It's necessary to compare different adaptation methods in terms of accent level. We manually categorized the 10 speakers into 3 accented groups: slight, medium and heavy according to their performances on the baseline AM in Table~\ref{tab:baseline}. According to the accent level, results are summarized in Figure~\ref{fig:accent_slight} (slight), Figure~\ref{fig:accent_moderate} (medium) and Figure~\ref{fig:accent_thick} (heavy). From Figure~\ref{fig:accent_slight}, we can see that LHUC performs consistently the best for the adaptation on slight-accent speakers, while KLD and RSI are not stable. We believe that this is because the baseline model is trained using data mostly from Mandarin speakers with standard accent and the baseline model itself is robust enough; in this case, direct update on the network parameters may be harmful. Observing Figure~\ref{fig:accent_moderate}, for medium-accent speakers, we can see that KLD and LHUC can get comparable performances with much lower CER than LIN. RSI is still not stable and over-fitting happens when a large adaptation data set is used. If memory footprint is a major consideration, we suggest to use LHUC as its has a small set of adapted parameters for each speaker; otherwise LHUC and KLD can be both  considered for medium-accent speakers. As shown in Figure~\ref{fig:accent_thick}, for heavy-accent speakers, KLD can get absolutely the best performance among the three methods, followed by LHUC, while LIN still performs the worst. We believe that KLD's superior performance is because the posterior distribution of the heavy-accent speech is far away from that of the unaccented speech; in this case, directly updating the network parameters or dragging the two distributions closer, is the most effective means. This also clarifies why RSI is better than LHUC and why we cannot observe over-fitting in this condition.

\section{Conclusions}\label{section:conclusions}

In this work, we have systematically compared the performance of three widely-used speaker adaptation methods on a challenging dataset with accented speakers. We show that i-vector based SAT-DNN AM is already strong enough to slight-accent speakers but performs badly to medium- and heavy-accent speakers. By using LIN,  KLD, LHUC, we can further improve the speech recognition performance not only for medium- and heavy-accent speakers, but also for slight-accent speakers. Moreover, the experimental results show that, in general, KLD and LHUC consistently outperform LIN and KLD demonstrates the best performance. The combination of different methods cannot bring salient improvements. For the adaptation on slight-accent speakers, LHUC is preferred with consistent improvement, while KLD and RSI are not stable. For medium-accent speakers, KLD and LHUC can get comparable performances with much lower CER than LIN. For heavy-accent speakers, KLD can get absolutely the best performance, followed by LHUC, while LIN still performs the worst.

\section{Acknowledgements}\label{section:acknowledgements}

The authors would like to thank Jian Li, Mengfei Wu and Yongqing Wang for their supports on this work. The research work is supported by the National Key Research and Development Program of China (Grant No.2017YFB1002102) and the National Natural Science Foundation of China (Grant No.61571363).

\bibliographystyle{IEEEtran}

\bibliography{interspeech2018-Adapt}

\begin{thebibliography}{10}
\providecommand{\url}[1]{#1}
\csname url@samestyle\endcsname
\providecommand{\newblock}{\relax}
\providecommand{\bibinfo}[2]{#2}
\providecommand{\BIBentrySTDinterwordspacing}{\spaceskip=0pt\relax}
\providecommand{\BIBentryALTinterwordstretchfactor}{4}
\providecommand{\BIBentryALTinterwordspacing}{\spaceskip=\fontdimen2\font plus
\BIBentryALTinterwordstretchfactor\fontdimen3\font minus
  \fontdimen4\font\relax}
\providecommand{\BIBforeignlanguage}[2]{{%
\expandafter\ifx\csname l@#1\endcsname\relax
\typeout{** WARNING: IEEEtran.bst: No hyphenation pattern has been}%
\typeout{** loaded for the language `#1'. Using the pattern for}%
\typeout{** the default language instead.}%
\else
\language=\csname l@#1\endcsname
\fi
#2}}
\providecommand{\BIBdecl}{\relax}
\BIBdecl

\bibitem{dahl2012context}
G.~E. Dahl, D.~Yu, L.~Deng, and A.~Acero, ``Context-dependent pre-trained deep
  neural networks for large-vocabulary speech recognition,'' \emph{IEEE
  Transactions on audio, speech, and language processing}, vol.~20, no.~1, pp.
  30--42, 2012.

\bibitem{hinton2012deep}
G.~Hinton, L.~Deng, D.~Yu, G.~E. Dahl, A.~R. Mohamed, N.~Jaitly, A.~Senior,
  V.~Vanhoucke, P.~Nguyen, T.~N. Sainath \emph{et~al.}, ``Deep neural networks
  for acoustic modeling in speech recognition: The shared views of four
  research groups,'' \emph{IEEE Signal Processing Magazine}, vol.~29, no.~6,
  pp. 82--97, 2012.

\bibitem{abdel2012applying}
O.~Abdel-Hamid, A.~R. Mohamed, H.~Jiang, and G.~Penn, ``Applying convolutional
  neural networks concepts to hybrid nn-hmm model for speech recognition,'' in
  \emph{IEEE International Conference on Acoustics, Speech and Signal
  Processing}, 2012, pp. 4277--4280.

\bibitem{abdel2013exploring}
O.~Abdel-Hamid, L.~Deng, and D.~Yu, ``Exploring convolutional neural network
  structures and optimization techniques for speech recognition.'' in
  \emph{Interspeech}, vol. 2013, 2013, pp. 1173--5.

\bibitem{peddinti2015time}
V.~Peddinti, D.~Povey, and S.~Khudanpur, ``A time delay neural network
  architecture for efficient modeling of long temporal contexts,'' in
  \emph{Sixteenth Annual Conference of the International Speech Communication
  Association}, 2015.

\bibitem{sak2014long}
H.~Sak, A.~Senior, and F.~Beaufays, ``Long short-term memory based recurrent
  neural network architectures for large vocabulary speech recognition,''
  \emph{arXiv preprint arXiv:1402.1128}, 2014.

\bibitem{sak2015fast}
H.~Sak, A.~Senior, K.~Rao, and F.~Beaufays, ``Fast and accurate recurrent
  neural network acoustic models for speech recognition,'' \emph{arXiv preprint
  arXiv:1507.06947}, 2015.

\bibitem{zhang2016highway}
Y.~Zhang, G.~Chen, D.~Yu, K.~Yaco, S.~Khudanpur, and J.~Glass, ``Highway long
  short-term memory rnns for distant speech recognition,'' in \emph{Acoustics,
  Speech and Signal Processing (ICASSP), 2016 IEEE International Conference
  on}.\hskip 1em plus 0.5em minus 0.4em\relax IEEE, 2016, pp. 5755--5759.

\bibitem{zhang2015feedforward}
S.~Zhang, C.~Liu, H.~Jiang, S.~Wei, L.~Dai, and Y.~Hu, ``Feedforward sequential
  memory networks: A new structure to learn long-term dependency,'' \emph{arXiv
  preprint arXiv:1512.08301}, 2015.

\bibitem{sainath2015convolutional}
T.~N. Sainath, O.~Vinyals, A.~Senior, and H.~Sak, ``Convolutional, long
  short-term memory, fully connected deep neural networks,'' in
  \emph{Acoustics, Speech and Signal Processing (ICASSP), 2015 IEEE
  International Conference on}.\hskip 1em plus 0.5em minus 0.4em\relax IEEE,
  2015, pp. 4580--4584.

\bibitem{Woodland2001Speaker}
P.~C. Woodland, ``Speaker adaptation for continuous density hmms: A review,''
  2001.

\bibitem{Gauvain1994Maximum}
J.~L. Gauvain and C.~H. Lee, ``Maximum a posteriori estimation for multivariate
  gaussian mixture observations of markov chains,'' \emph{IEEE Transactions on
  Speech and Audio Processing}, vol.~2, no.~2, pp. 291--298, 1994.

\bibitem{Legetter1995Maximum}
C.~J. Legetter and P.~C. Woodland, ``Maximum likelihood linear regression
  speaker adaptation of continuous density hmms,'' \emph{Computer Speech and
  Language}, 1995.

\bibitem{Digalakis1995Speaker}
V.~V. Digalakis, D.~Rtischev, and L.~G. Neumeyer, ``Speaker adaptation using
  constrained estimation of gaussian mixtures,'' \emph{IEEE Transactions on
  Speech and Audio Processing}, vol.~3, no.~5, pp. 357--366, 1995.

\bibitem{Gales1998Maximum}
M.~J.~F. Gales, ``Maximum likelihood linear transformations for hmm-based
  speech recognition,'' \emph{Computer Speech and Language}, vol.~12, no.~2, p.
  75–98, 1998.

\bibitem{Gales2000Cluster}
------, ``Cluster adaptive training of hidden markov models,'' \emph{Speech and
  Audio Processing IEEE Transactions on}, vol.~8, no.~4, pp. 417--428, 2000.

\bibitem{Kuhn2000Rapid}
R.~Kuhn, J.~C. Junqua, P.~Nguyen, and N.~Niedzielski, ``Rapid speaker
  adaptation in eigenvoice space,'' \emph{IEEE Trans Speech Audio Proc},
  vol.~8, no.~6, pp. 695--707, 2000.

\bibitem{Uebel1999An}
L.~F. Uebel and P.~C. Woodland, ``An investigation into vocal tract length
  normalisation,'' in \emph{European Conference on Speech Communication and
  Technology, Eurospeech 1999, Budapest, Hungary, September}, 1999.

\bibitem{neto1995speaker}
J.~Neto, L.~Almeida, M.~Hochberg, C.~Martins, L.~Nunes, S.~Renals, and
  T.~Robinson, ``Speaker-adaptation for hybrid hmm-ann continuous speech
  recognition system,'' in \emph{Fourth European Conference on Speech
  Communication and Technology}, 1995.

\bibitem{li2010comparison}
B.~Li and K.~C. Sim, ``Comparison of discriminative input and output
  transformations for speaker adaptation in the hybrid nn/hmm systems,'' in
  \emph{Eleventh Annual Conference of the International Speech Communication
  Association}, 2010.

\bibitem{gemello2007linear}
R.~Gemello, F.~Mana, S.~Scanzio, P.~Laface, and R.~De~Mori, ``Linear hidden
  transformations for adaptation of hybrid ann/hmm models,'' \emph{Speech
  Communication}, vol.~49, no. 10-11, pp. 827--835, 2007.

\bibitem{swietojanski2014learning}
P.~Swietojanski and S.~Renals, ``Learning hidden unit contributions for
  unsupervised speaker adaptation of neural network acoustic models,'' in
  \emph{Spoken Language Technology Workshop (SLT), 2014 IEEE}.\hskip 1em plus
  0.5em minus 0.4em\relax IEEE, 2014, pp. 171--176.

\bibitem{saon2013speaker}
G.~Saon, H.~Soltau, D.~Nahamoo, and M.~Picheny, ``Speaker adaptation of neural
  network acoustic models using i-vectors.'' in \emph{ASRU}, 2013, pp. 55--59.

\bibitem{miao2015speaker}
Y.~Miao, H.~Zhang, and F.~Metze, ``Speaker adaptive training of deep neural
  network acoustic models using i-vectors,'' \emph{IEEE/ACM Transactions on
  Audio, Speech and Language Processing (TASLP)}, vol.~23, no.~11, pp.
  1938--1949, 2015.

\bibitem{abdel2013fast}
O.~Abdel-Hamid and H.~Jiang, ``Fast speaker adaptation of hybrid nn/hmm model
  for speech recognition based on discriminative learning of speaker code,'' in
  \emph{Acoustics, Speech and Signal Processing (ICASSP), 2013 IEEE
  International Conference on}.\hskip 1em plus 0.5em minus 0.4em\relax IEEE,
  2013, pp. 7942--7946.

\bibitem{yu2013kl}
D.~Yu, K.~Yao, H.~Su, G.~Li, and F.~Seide, ``Kl-divergence regularized deep
  neural network adaptation for improved large vocabulary speech recognition,''
  in \emph{Acoustics, Speech and Signal Processing (ICASSP), 2013 IEEE
  International Conference on}.\hskip 1em plus 0.5em minus 0.4em\relax IEEE,
  2013, pp. 7893--7897.

\bibitem{senior2014improving}
A.~Senior and I.~Lopez-Moreno, ``Improving dnn speaker independence with
  i-vector inputs,'' in \emph{Acoustics, Speech and Signal Processing (ICASSP),
  2014 IEEE International Conference on}.\hskip 1em plus 0.5em minus
  0.4em\relax IEEE, 2014, pp. 225--229.

\bibitem{huang2017bayesian}
Z.~Huang, S.~M. Siniscalchi, and C.-H. Lee, ``Bayesian unsupervised batch and
  online speaker adaptation of activation function parameters in deep models
  for automatic speech recognition,'' \emph{IEEE/ACM Transactions on Audio,
  Speech, and Language Processing}, vol.~25, no.~1, pp. 64--75, 2017.

\bibitem{saon2015ibm}
G.~Saon, H.-K.~J. Kuo, S.~Rennie, and M.~Picheny, ``The ibm 2015 english
  conversational telephone speech recognition system,'' \emph{arXiv preprint
  arXiv:1505.05899}, 2015.

\bibitem{povey2016purely}
D.~Povey, V.~Peddinti, D.~Galvez, P.~Ghahremani, V.~Manohar, X.~Na, Y.~Wang,
  and S.~Khudanpur, ``Purely sequence-trained neural networks for asr based on
  lattice-free mmi.'' in \emph{Interspeech}, 2016, pp. 2751--2755.

\bibitem{xiong2017microsoft}
W.~Xiong, J.~Droppo, X.~Huang, F.~Seide, M.~Seltzer, A.~Stolcke, D.~Yu, and
  G.~Zweig, ``The microsoft 2016 conversational speech recognition system,'' in
  \emph{Acoustics, Speech and Signal Processing (ICASSP), 2017 IEEE
  International Conference on}.\hskip 1em plus 0.5em minus 0.4em\relax IEEE,
  2017, pp. 5255--5259.

\bibitem{povey2011kaldi}
D.~Povey, A.~Ghoshal, G.~Boulianne, L.~Burget, O.~Glembek, N.~Goel,
  M.~Hannemann, P.~Motlicek, Y.~Qian, P.~Schwarz \emph{et~al.}, ``The kaldi
  speech recognition toolkit,'' in \emph{IEEE 2011 workshop on automatic speech
  recognition and understanding}, no. EPFL-CONF-192584.\hskip 1em plus 0.5em
  minus 0.4em\relax IEEE Signal Processing Society, 2011.

\end{thebibliography}

\end{spacing}
\end{document}